\documentclass[12pt,preprint]{aastex}

\slugcomment{to appear in the ApJ}

\shorttitle{@@@}
\shortauthors{@@@}

\begin{document}

\title{An XMM-$Newton$ and {\it Chandra} investigation of the nuclear
accretion in the Sombrero Galaxy (NGC~4594)}


\author{S. Pellegrini}
\affil{Astronomy Department, Bologna University, Italy}
\email{pellegrini@bo.astro.it}

\author{A. Baldi}

\author{G. Fabbiano}

\author{D.-W. Kim}
\affil{Harvard-Smithsonian Center for Astrophysics, Cambridge, MA, USA}
\email{abaldi, pepi, kim@head-cfa.harvard.edu}


\begin{abstract}
We present an analysis of the XMM-$Newton$ and $Chandra$ ACIS-S
observations of the LINER nucleus of the Sombrero galaxy and we
discuss possible explanations for its very sub-Eddington luminosity by
complementing the X-ray results with high angular resolution
observations in other bands. The X-ray investigation shows a hard
($\Gamma =1.89$) and moderately absorbed ($N_H=1.8\times 10^{21}$
cm$^{-2}$) nuclear source of $1.5\times 10^{40}$ erg s$^{-1}$ in the
2--10 keV band, surrounded by hot gas at a temperature of $\sim 0.6$
keV.  The bolometric nuclear luminosity is at least $\sim 200$ times
lower than expected if mass accreted on the supermassive black hole,
that $HST$ shows to reside at the center of this galaxy, at the rate
predicted by the spherical and adiabatic Bondi accretion theory and
with the high radiative efficiency of a standard accretion disc. The
low luminosity, coupled to the observed absence of Fe-K emission in
the nuclear spectrum, indicates that such a disc is not present. This
nucleus differs from bright unobscured AGNs also for the lack of high
flux variability and of prominent broad $H\alpha$ emission. However,
it is also too faint for the predictions of simple radiatively
inefficient accretion taking place at the Bondi rate; it could be too
radio bright, instead, for radiatively inefficient accretion that
includes strong mass outflows or convection.  This discrepancy could
be solved by the possible presence of nuclear radio jets. An
alternative explanation of the low luminosity, in place of radiative
inefficiency, could be unsteady accretion.

\end{abstract}



\section{Introduction}

Nearby galactic spheroids are believed to host supermassive black
holes \citep[SMBHs; e.g.,][]{kor95,ric98}. However, only low level
LINER activity \citep{ho97} or no sign of activity are almost the rule
for these systems. This radiative quiescence represents one of the
most intriguing aspects of SMBHs in the nearby Universe. The Sombrero
galaxy (NGC~4594) is well known to host a SMBH at its nucleus, whose
mass ($\sim 10^9 M_{\odot}$) has been established directly by $HST$
measurements \citep{kor96}.  The nucleus of Sombrero is also a LINER 2
\citep{hec80,ho97}, a compact and variable radio continuum source
\citep[e.g.][]{hum84,baj88} and a pointlike source in an $HST$ high
resolution image \citep{cra93}. Thanks to its proximity (d=9.4 Mpc,
Ajhar et al. 1997) and the availability of high angular resolution
observations, this nucleus represents an ideal case to study in detail
how the modality of accretion changes from high luminosity to low
luminosity AGNs.  The answer to this question is key to the knowledge
of the history of accretion in the Universe \citep[e.g.,][]{hai03}.

Sombrero was first detected in X-rays with {\it Einstein}. Although
its X-ray luminosity was included in a list of LINERs \citep{hal83},
subsequent more detailed work showed an extended source
\citep[see][]{for85,fab92}. A ROSAT HRI image detected a luminous,
possibly variable, pointlike source associated with the nucleus
\citep{fab97}, with a (0.1--2.4)~keV luminosity of $\sim 9 \times
10^{39}$ erg s$^{-1}$ (when rescaled for the distance adopted
here). ASCA and BeppoSAX spectra suggest the presence of a moderately
absorbed power law (of photon index $\Gamma \sim 1.9$), as well as a
softer thermal component of $kT\sim 0.6$ keV \citep{nic98,ter02,pel02}. An
upper limit of 150 eV and 260 eV was placed on the equivalent width
of an Fe-K line at 6.4 keV and 6.7 keV respectively.  A $Chandra$
snapshot exposure showed an image dominated by the nucleus
\citep{ho01} whose spectrum again was found consistent with a
moderately absorbed hard power law \citep{pel02}.

Comparison of the X-ray, H$\alpha$ and radio emission of the nuclear
source with those of bright AGNs suggests that it may be a low luminosity
version of the latter \citep{fab97,pel02}. 
However, the shape of the whole spectral energy distribution of this
nucleus compares well with those of other low luminosity AGNs
\citep{ho99}, all of which reveal the absence of a big blue bump.

If the nuclear X-ray emission is due to accretion onto the SMBH, 
its luminosity is extremely sub-Eddington
\citep[$L(2-10$ keV$) \sim 10^{-7} L_{Edd}$,][]{pel02}. 
This fact, coupled to the absence of an Fe-K detection, suggest that
this nucleus may be a good test case for radiatively inefficient
accretion \citep{ree82,nar95}. It is also interesting
to compare this nucleus to other recently studied sub-Eddington nuclei
of ``normal'' elliptical galaxies, such as IC~1459 \citep{fab03} and
NGC1399, NGC4636, NGC4472 \citep{low01}.

We report here a study of the nucleus of the Sombrero galaxy based on
deep XMM and {\it Chandra} observations. XMM's excellent sensitivity
at both soft and hard X-rays allows us to measure accurately the hard
power law component and set stringent limits on the Fe-K emission. The
high resolution archival $Chandra$ ACIS-S observation allows us to
study the circumnuclear region and constrain the amount of fuel
available to feed the nucleus. With both results we obtain a more
complete picture of the physical state of this X-ray faint LINER
nucleus.

\section{Observations and Data preparation} \label{dataprep}

Table~\ref{obslog} summarizes the observing log.  The Sombrero galaxy
was observed by XMM-$Newton$ during revolution 376 (PI: Fabbiano;
ObsID 0084030101) on December 28, 2001 for a total exposure time of
40~ks, and with {\em Chandra} ACIS-S (PI: Murray; ObsID 1586) on May
31, 2001 for a total of 18.7 ks.

The XMM-$Newton$ data were processed using the Science Analysis
Software ({\em XMM-SAS}) v5.3.  The event files were 'cleaned' of
background flares, caused by soft ($<$ a few 100 keV) protons hitting
the detector surface.  These flares were found analyzing the light
curves at energies greater than 10 keV (in order to avoid contribution
from real X-ray source variability) and setting a threshold for good
time intervals at 0.35 cts/s for each {\em MOS} unit and at 0.55 cts/s
for the {\em pn} unit.  Soft proton flares removal and a 
duration of the observation of about
half the scheduled exposure time resulted in a substantial
reduction of the useful integration time in the {\em pn}, for which we
obtained 17 ks. For the {\em MOS1} and the {\em MOS2} instead we
obtained 34.6 ks and 40.5 ks, respectively.

We used the archival {\it Chandra} observation in conjunction with
the XMM-$Newton$ data to derive information on the nuclear
surroundings on the small scale. The {\em Chandra} data were processed
using CIAO v2.3 data analysis software.

\section{Spectral analysis of the XMM nuclear data} \label{nucleus}

\begin{figure}
\epsscale{0.90}
\plotone{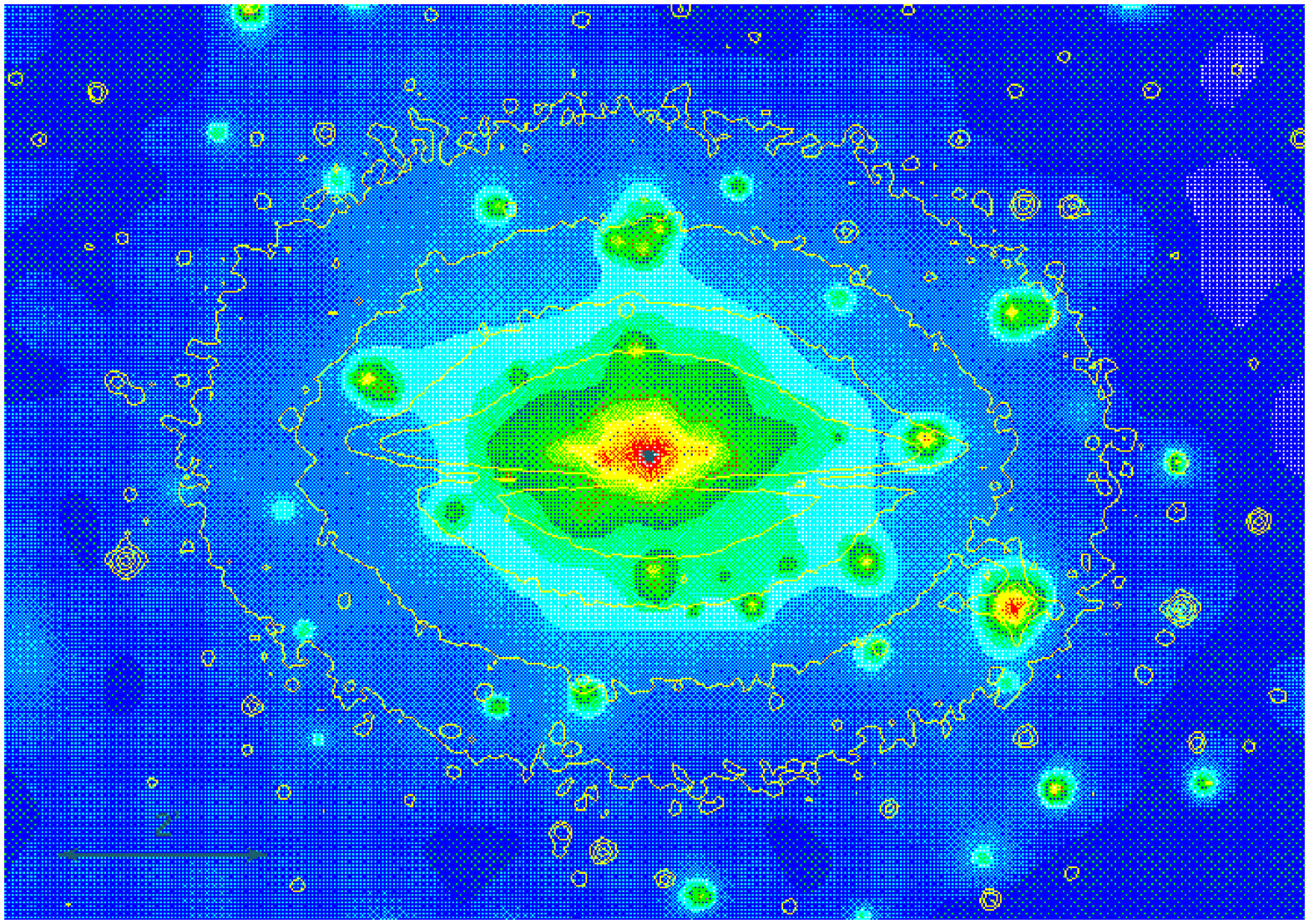}
\caption{Adaptively smoothed XMM-$Newton$ EPIC image of the Sombrero galaxy (0.5-2 keV band) with 
superimposed optical contours from the DSS. North is up and East to the left.
\label{xmmfig}}
\end{figure}

An EPIC  XMM-$Newton$ view of the Sombrero galaxy is shown in
Figure~\ref{xmmfig}. The image shows a prominent pointlike nuclear
source surrounded by diffuse emission and several
off-nuclear pointlike sources (most likely luminous galaxian X-ray
binaries).

\begin{figure}
\epsscale{0.90}
\plotone{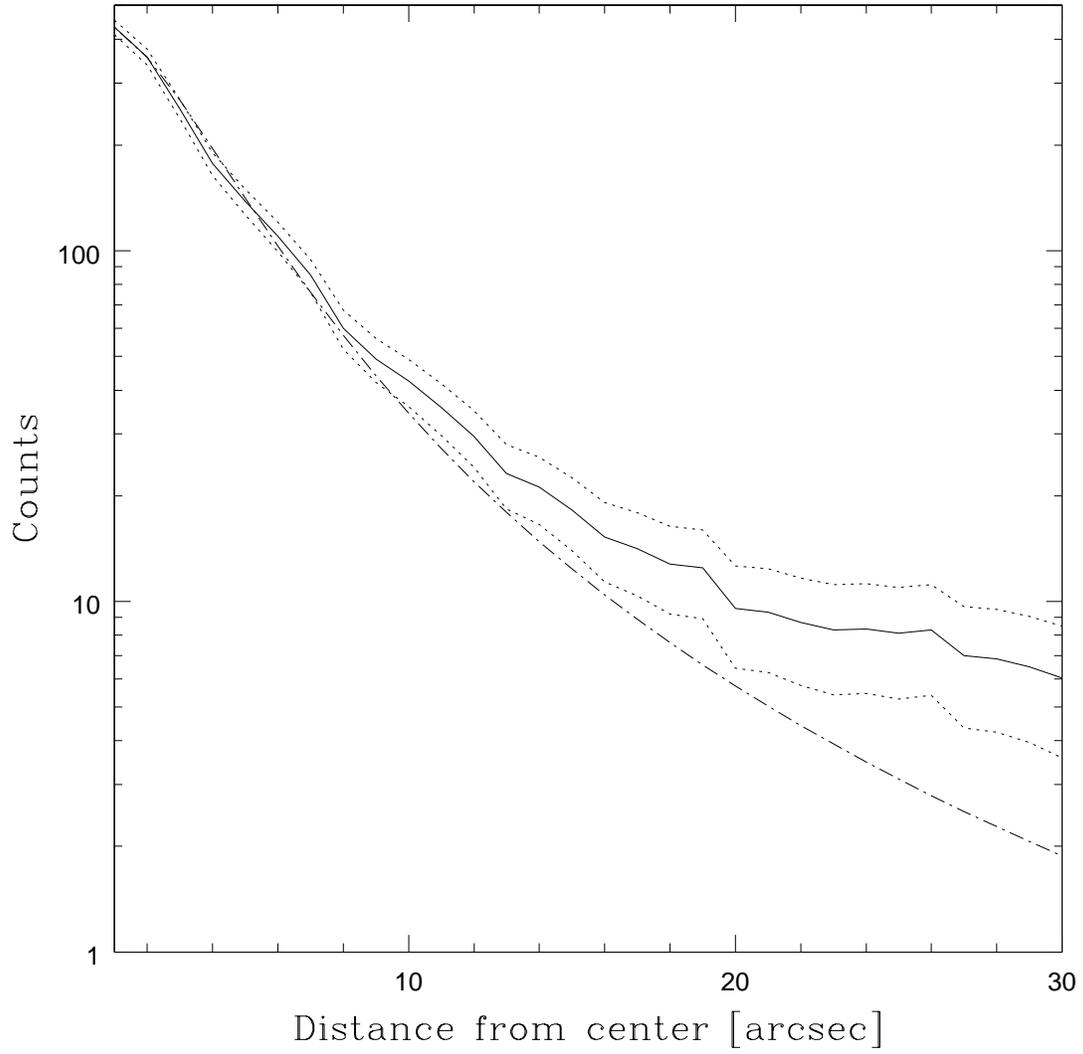}
\caption{0.5-2 keV band radial profile of the nucleus of Sombrero in EPIC MOS1 instrument 
compared with the instrumental PSF (dash-dotted line).
The solid line represents the measured counts while dotted lines are the errors on the counts.
\label{radial}}
\end{figure}

To investigate the spatial properties of the nuclear emission, we have
compared the radial profile of the nuclear source with the EPIC point
response function (PSF; Figure~\ref{radial}) normalized to the
detected counts in the central $1^{\prime\prime}$ bin around the
nucleus. The radial profile agrees well with the PSF out to a radius
of $\sim 7-8^{\prime\prime}$. At larger radii, the contribution of
non-nuclear emission becomes important. Given the presence of both
pointlike and extended emission, we have performed spectral analyses
on the counts extracted from two circles of different radii
($20^{\prime\prime}$ and $7^{\prime\prime}$) centered on the nucleus,
and compared the results to assess the uncertainty on the spectral
parameters of the nuclear source resulting from galaxian
contamination.  The $20^{\prime\prime}$ radius corresponds to
$\sim0.9$ kpc at the distance of the galaxy 
and to an encircled energy fraction (EEF) of 0.75; the
$7^{\prime\prime}$ radius corresponds to a physical size of $\sim0.3$
kpc and to an EEF of 0.42.  In both cases, the background was
extracted from a source free circular region outside the optical
galaxy having a radius of 80 arcsec.  We used both EPIC {\em MOS} and
{\em pn } data, coadding MOS1 and MOS2 spectra, to optimize the
signal-to-noise ratio of the data.  The latest versions of the response
matrices (v6.2) were used for the fit. The spectra were rebinned in
order to have at least 20 counts for each energy bin and we used XSPEC
for the spectral analysis. We fitted together the {\em
MOS} and {\em pn} data, leaving their relative normalization free to
vary.

Tables~\ref{specfit1} and \ref{specfit2} summarize the results of the
spectral analysis.  All the errors are quoted at 90\% for two
interesting parameters.  We used two different models: a simple
absorbed power law model (XSPEC model: $wabs(zwabs(pow))$, where
$wabs$ is fixed at the Galactic absorbing column of $N_H=3.7\times
10^{20}$ cm$^{-2}$, \citep{sta92} ) and a composite power law plus
thermal model ($wabs(mekal+zwabs(pow))$, with abundance fixed at 0.5
solar), in order to model the emission from a possible extended
circumnuclear hot interstellar medium. 

The simple power law fit gives a $\chi^2$ of 556.9 with 527 degrees of
freedom (corresponding to a probability of 17.8\%) for the
$20^{\prime\prime}$ data, and a $\chi^2$ of 330.2 with 320 degrees of
freedom (probability of 33.6\%) for the $7^{\prime\prime}$ data. The
best fit photon index is the same within the errors for the two
spectra ($\Gamma=1.89\pm 0.03$ and $\Gamma =1.88^{+0.05}_{-0.04}$
respectively). It is consistent with that obtained from BeppoSAX data
and is somewhat higher than, although still marginally consistent
with, that found from a spectral analysis of the nucleus with the
snapshot ACIS-S exposure \citep{pel02}.  The intrinsic absorbing
column density is only moderate; it increases for the smaller
extraction radius [going from $(1.3\pm 0.1)$ to $(1.8\pm 0.1) \, \times
10^{21} $ cm$^{-2}$], which suggests contamination by a soft possibly
thermal component outside the $7^{\prime\prime}$ radius. The $N_H$
from the $7^{\prime\prime}$ data is consistent with that from the
analysis of the snapshot ACIS-S exposure and marginally higher than
the BeppoSAX value, which can easily be due to the difference in the
extraction radius.

\begin{figure}
\rotatebox{-90}{
\epsscale{0.90}
\plottwo{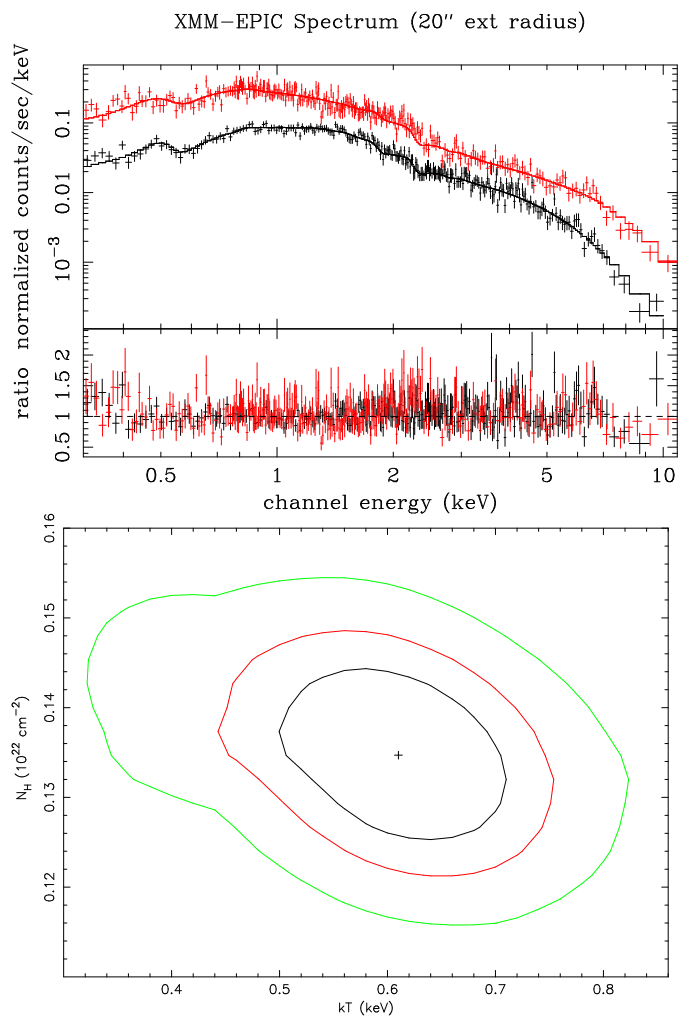}{f3b.eps}}
\caption{{\em Upper panel:} Sombrero nuclear EPIC MOS (black) and pn (red) spectra and fit residuals for an absorbed 
power law model plus thermal component, obtained from a 20 arcsec extraction radius. {\em Lower panel:} Confidence
contours at 68\%, 90\% and 99\% for the best-fit temperature $kT$ and intrinsic absorbing column
density $N_H$.
\label{spectrum20}}
\end{figure}

\begin{figure}
\rotatebox{-90}{
\epsscale{0.90}
\plottwo{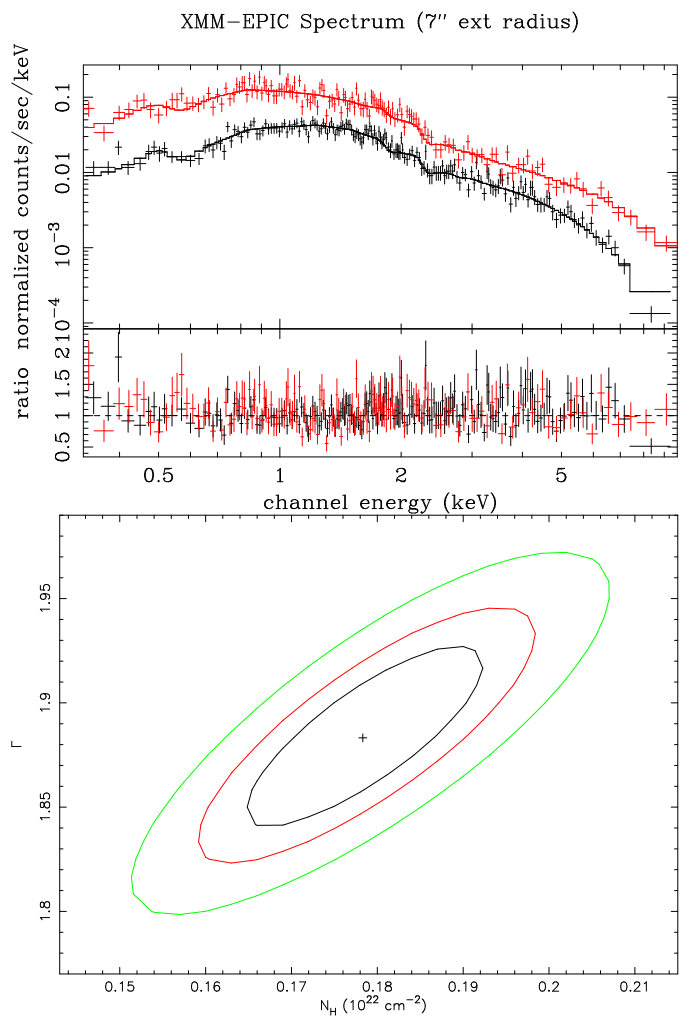}{f4b.eps}}
\caption{{\em Upper panel:} Sombrero nuclear EPIC MOS (black) and pn (red) spectra and fit residuals for a simple absorbed 
power law model, obtained from a 7 arcsec extraction radius. {\em Lower panel:} Confidence
contours at 68\%, 90\% and 99\% for the best-fit intrinsic absorbing column density $N_H$ and photon
index $\Gamma$.
\label{spectrum7}}
\end{figure}

The power law + thermal model fit gives a better representation of the
$20^{\prime\prime}$ spectrum, with a $\chi^2$ of 537.3 for 525 degrees
of freedom (probability of 34.6\%). An F-test demonstrates that this
improvement is significant (F-test significance $\sim10^{-4}$).  We
find no improvement instead for the $7^{\prime\prime}$ spectrum, where we
obtain a $\chi^2$ of 326 for 318 degrees of freedom (probability of
36.6\%).  In both cases, the intrinsic N$_H$ and the photon index are
consistent with those obtained from the simple power law fit.
The best fit temperature of the thermal component is $kT=0.61$
keV. Not surprisingly, the $20^{\prime\prime}$ fit returns a larger
value for the luminosity of this component.  The $20^{\prime\prime}$
and $7^{\prime\prime}$ spectra, their modelings and fit residuals, and
the confidence contours for the intrinsic $N_H$, $kT$ and $\Gamma$ are
shown in Figure~\ref{spectrum20} and \ref{spectrum7}.

Adding to the absorbed power law model an iron K$\alpha$ emission line
at 6.4 keV, for the $7^{\prime\prime}$ spectrum, does not improve the
fit. We can set an upper limit on the equivalent width (EW) of this line of
$145$ eV, for a narrow line, and of 296 eV for a broad Gaussian
line with line width $\sigma =0.43$ keV (as found on average
for Seyfert 1's by \citet{nan97b}).

We have not used the 18.7 ks exposure ACIS-S data for the spectral
analysis of the nucleus because they are strongly affected by pile-up.

\section{Mass accretion rate} \label{bondi}

\begin{figure}
\epsscale{0.90}
\plotone{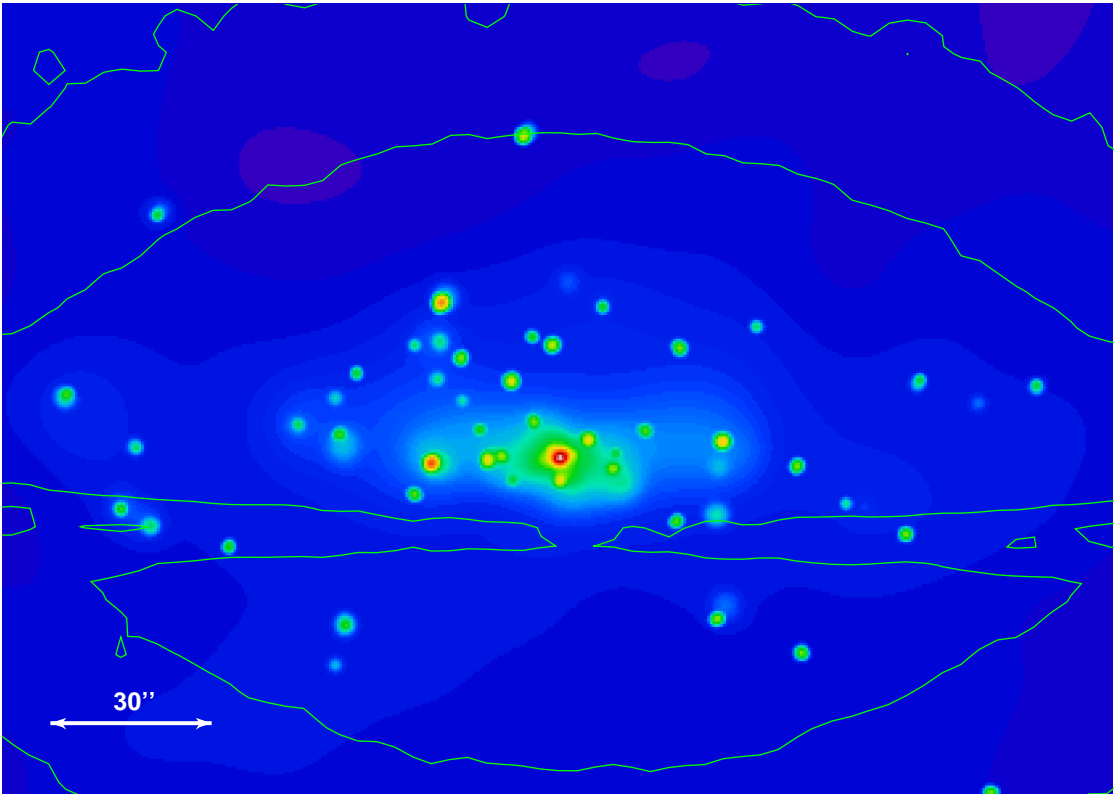}
\caption{Adaptively smoothed $Chandra$ ACIS-S image (0.3-10 keV band) 
of the Sombrero galaxy with superimposed optical contours from the DSS.
North is up and East to the left.
\label{imachandra}}
\end{figure}

While pileup makes the interpretation of {\it Chandra} data ambiguous
for the nuclear source, thanks to the very sharp {\it Chandra} PSF
these data provide unique information on the circumnuclear region, as
near as $2^{\prime\prime}$ from the nucleus, corresponding to $<
100$~pc at the distance of Sombrero.  Figure~\ref{imachandra} shows a
view of the central regions of this galaxy resulting from the 18.7 ks
ACIS-S exposure; the image has been obtained with adaptive smoothing,
i.e., after applying a tool that smoothes a two-dimensional image with a
circular Gaussian kernel of varying size (here, from 1 to 20 pixels;
see http://cxc.harvard.edu/ciao/ahelp/csmooth.html).

\begin{figure}
\epsscale{1.10}
\plottwo{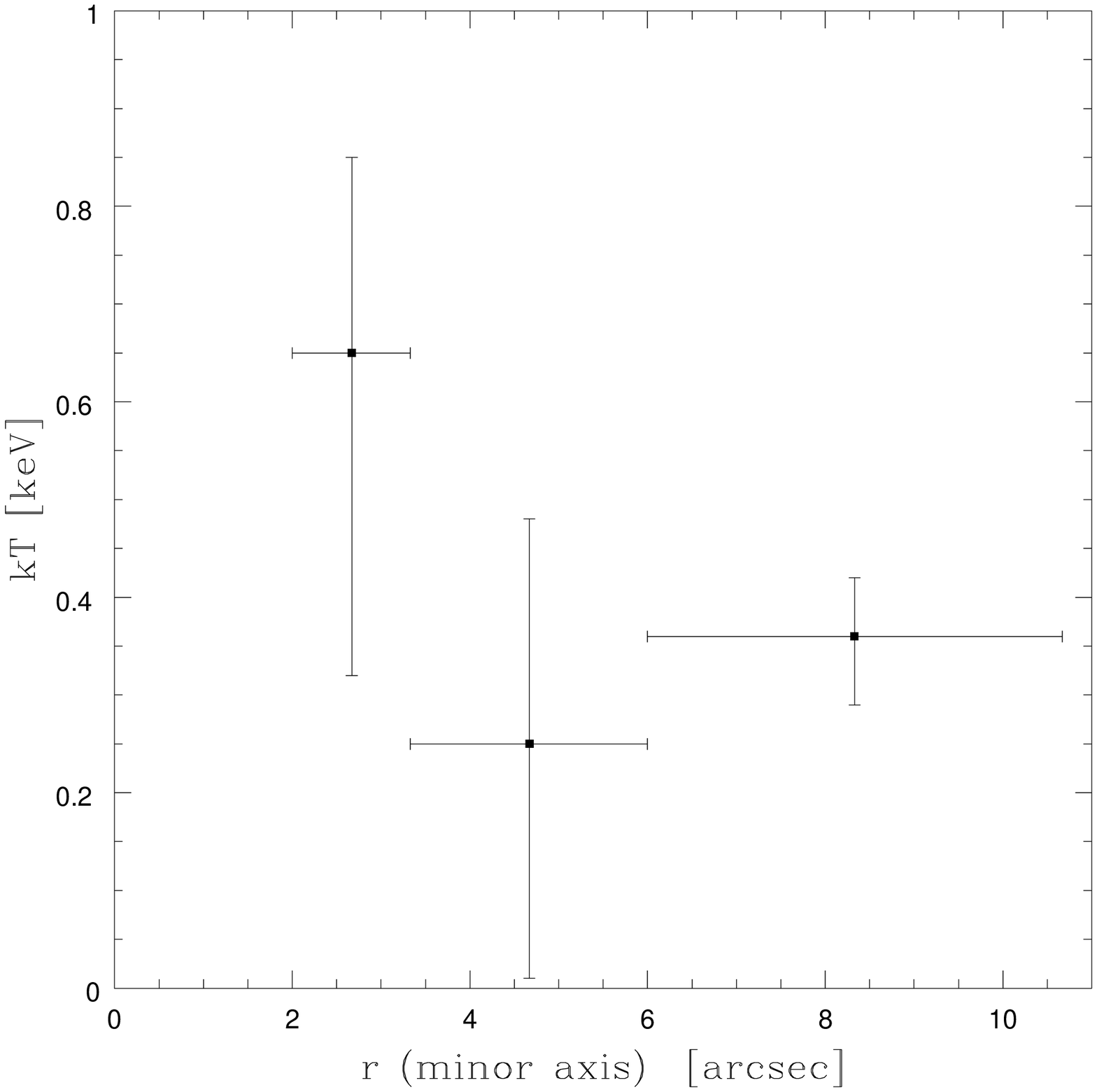}{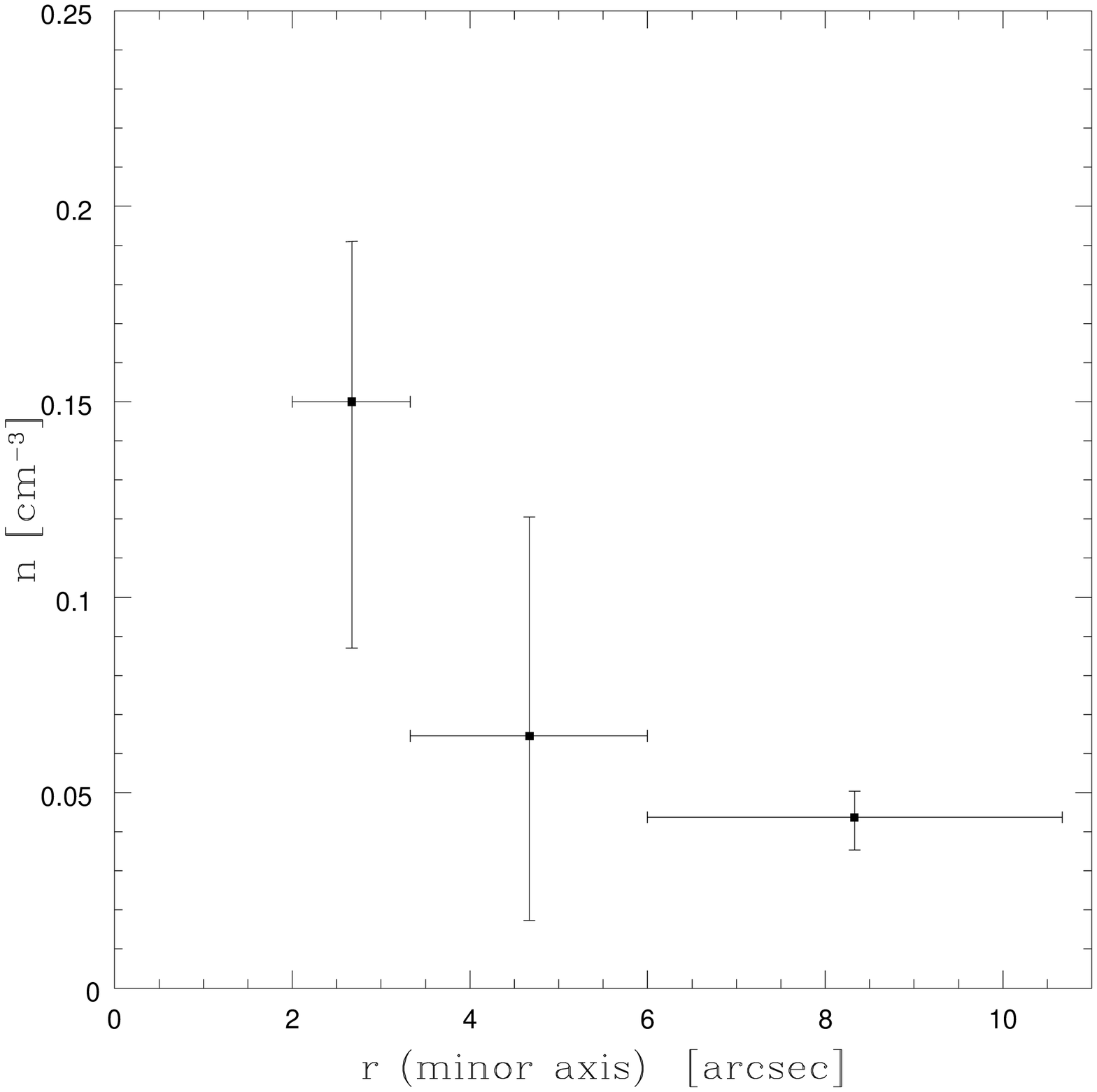}
\caption{Temperature profile (left panel) and density profile (right panel) 
for the Sombrero galaxy. Errorbars indicate the $ 90$\% confidence 
interval.  }
\label{profiles}
\end{figure}

The high resolution ACIS-S data allow us to derive the density and
temperature profile of the interstellar matter in the surroundings of
the nuclear SMBH by using a deprojection technique implemented within
XSPEC.  Spectra extracted from three elliptical annuli centered on the
nucleus were compared with the spectra expected from the superposition
along the line of sight of the emission coming from the corresponding
ellipsoidal shells.  The elliptical annuli have a major/minor axis
ratio of 1.5, equal to that of the Sombrero's bulge.  To avoid
contamination from the nucleus, we chose as inner boundary of the
first annulus a $2^{\prime\prime}$ circle; its outer boundary is an
ellipse with semiminor axis of $3^{\prime\prime}\hskip -0.1truecm .33$
and semimajor axis of $5^{\prime\prime}$. The outer boundary of the
second annulus has a semiminor axis of $6^{\prime\prime}$ and a
semimajor axis of $9^{\prime\prime}$.  The third annulus has as outer
boundary an ellipse with semiminor axis of $10^{\prime\prime}\hskip
-0.1truecm .67$ and a semimajor axis of $16^{\prime\prime}$.  From
these annuli we subtracted clearly visible point sources. Assuming
that the emission comes from hot gas plus residual contamination from
unresolved binaries \citep[see, e.g.,][]{kim03}, we obtained
deprojected density and temperature profiles by fitting together the
spectra of the 3 annuli with the XSPEC model {\it
projct*wabs(raymond+powerlaw)}.  The power law index was fixed at
$\Gamma=1.9$, $wabs$ was fixed at the Galactic value and the
temperature and normalizations were free parameters.  The resulting
temperature and density profiles are shown in Figure~\ref{profiles}.

The above calculation was next used to derive an estimate of the mass
supply rate for accretion on the SMBH, by applying the Bondi (1952)
theory of steady, spherical and adiabatic accretion. This requires $T$
and $n$ at ``infinity'', in practice near the accretion radius (that
in the case of Sombrero is $r_A=GM_{BH}/c_s^2\sim 30$ pc, with $c_s$
the sound speed).  The inner radius of the innermost annulus ($\sim
90$ pc) is a factor of $\sim 3$ larger than $r_A$.  Moreover, the
temperature and density profiles show a sharp positive gradient toward
smaller radii. Therefore, in order to have an estimate of the Bondi
mass accretion rate ($ \dot{M}_{Bondi}$), an extrapolation of the $kT$
and $n$ values at $r_A$ is required. Since the uncertainties on these
values are large (Fig.~\ref{profiles}), we derived for $
\dot{M}_{Bondi}$ the maximum range allowed by the uncertainties, via a
linear extrapolation of the outer $T,n$ data that also considers their
errorbars.  From the formula
$$
\dot{M}_{Bondi} = 6.2 \times 10^{23} \;
M_9^2\;\;T_{0.8}^{-3/2} \;n_{0.15} \; \hbox{$\thinspace 
g\; s^{-1}$} 
$$
where $M_9$ is the SMBH mass in units of $10^9\; M_{\odot}$ \citep[for
Sombrero $M_{BH}\simeq10^9\; M_{\odot}$,][]{kor96}, $T_{0.8}$ is the
temperature in units of 0.8 keV and $n_{0.15}$ is the density in units
of 0.15 cm$^{-3}$ [see, e.g., eq.  (3) of \citet{dim03} ], we found that
$\dot{M}_{Bondi,min}\simeq 0.008 M_{\odot}$ yr$^{-1}$ and
$\dot{M}_{Bondi,max}\simeq 0.067 M_{\odot}$ yr$^{-1}$.
The best fit
temperature and density values of the innermost annulus
($kT=0.65$ keV; $n=0.15$ cm$^{-3}$) give 
$\dot{M}_{Bondi}\simeq 0.013 M_{\odot}$ yr$^{-1}$.

\section{Discussion}

We discuss here possible scenarios to explain the very sub-Eddington
luminosity of the Sombrero's nucleus, using the results of our
analysis of high quality X-ray data and information coming from high
angular resolution observations in other bands.

\subsection {Heavily obscured emission?}

\begin{figure}
\epsscale{0.9}
\plotone{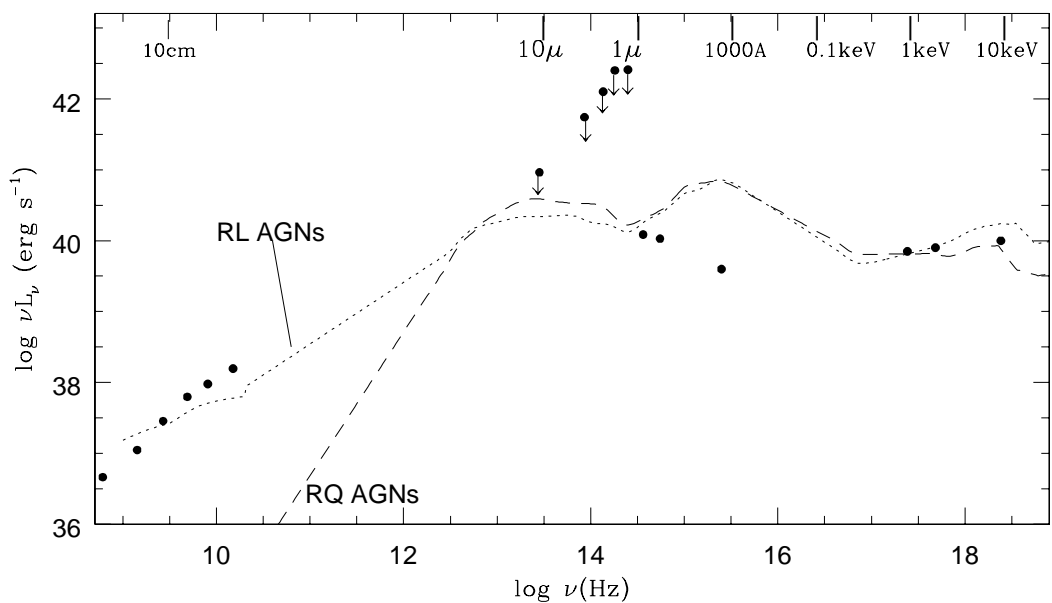}
\caption{The spectral energy distribution of the nucleus of Sombrero
from radio to X-rays.  Radio data with subarcsecond resolution are
from Hummel et al. 1984, IR data with $5^{\prime\prime}$ resolution
(plotted as upper limits) are from Willner et al. 1985, optical and UV
data with subarcsecond resolution are from the $HST$ WFPC2 images and
FOS spectrum (Ho 1999), the X-ray data are from this paper. Lines show the 
median distribution observed for low redshift
radio quiet and radio loud AGNs (Elvis et al. 1994), renormalized to
match the X-ray data.}
\label{sed}
\end{figure}

The bolometric luminosity of the Sombrero's nucleus is
$L_{bol}\sim 2.5\times 10^{41}$ erg s$^{-1}$ [obtained by integrating the
spectral energy distribution (SED) in Fig.~\ref{sed}], 
that is $2\times 10^{-6} L_{Edd}$. This
very sub-Eddington nuclear luminosity could be explained by Compton
thick material surrounding the nucleus and heavily absorbing the X-ray
emission, as in Seyfert~2 galaxies.  However, as observed in similar
nuclei \citep[e.g., IC~1459,][]{fab03}, though not with the high
XMM's sensitivity that we can exploit here, the spectral
characteristics of the emission are not consistent with this picture.
There is no sign of heavy obscuration and the small amount
of absorption needed for modeling the spectrum could be explained
by intervening dust: an $HST$ V--I image shows a dusty nuclear
environment \citep{pog00}, for which \citet{ems00} derive $A_V$=0.5
mag (that corresponds to an intrinsic $N_{H}=8.3\times 10^{20}$ cm$^{-2}$,
for the Galactic extinction law).  Moreover, the upper limit on the
equivalent width of Fe-K emission at 6.4 keV (EW$<145$ eV, Sect. 3) rules
out the presence of a strong 6.4 keV iron fluorescence line, of
EW$\ga 1$ keV, that is expected in the Compton thick scenario
\citep[as in NGC1068; e.g.,][]{mat97}.
Absence of high extinction is also supported by the $HST$ FOS spectra in the 
UV band \citep{nic98,ma98}.

\subsection{Downsized AGN?}

This nucleus shows some similarities with the observed emission
properties of luminous AGNs: an X-ray/radio luminosity ratio
comparable with the extrapolation of the bright radio galaxy
correlation \citep{fab84}; a spectral power law consistent with that
of Seyfert 1 galaxies \citep{nan97a}; 2--10 keV and $H\alpha$
luminosities consistent with the low luminosity extension of the
$L_X-L_{H\alpha}$ correlation found for powerful Seyfert 1 nuclei and
quasars \citep{war88,ho01}. 

However, its bolometric luminosity is well below what we would expect
from efficient accretion of the available fuel. The SMBH is surrounded
by hot gas whose density and temperature can be measured close to the
accretion radius. If at very small radius this gas joins an accretion
disc with a standard radiative efficiency $\eta \sim 0.1$, as in
brighter AGNs, it should produce a luminosity:
$$
L_{acc}=\eta \dot{M}_{Bondi}
c^2 \simeq (4.5-38) \times 10^{43} \hbox { erg s$^{-1}$}
$$
for the range of $\dot{M}_{Bondi}$ at the accretion radius estimated
in Sect. 4.  This is at least $\sim 200$ times higher than the
observed $L_{bol}$, which represents a strong argument against a thin
disc in this nucleus. An additional argument could be that its SED
(Fig.~\ref{sed}) lacks the `blue bump' (UV/blue excess) observed in
luminous AGNs and attributed to the accretion disc \citep{shi78}. This
lack is a feature common to other low luminosity AGNs \citep{ho99}.
However, the spectral energy distribution of a multicolor black body
accretion disc \citep[e.g.,][] {fra92}, with a mass accretion rate in
the range estimated in Sect. 4 for $\dot M_{Bondi}$ and $M_{BH}=10^9$
M$_{\odot}$, shows a peak at a frequency of $\sim (1.8-3.0)\times
10^{14}$ Hz, which is redder than in the B and U bands.

Besides the very low luminosity, inconsistent with efficient disc
accretion, there are other properties in which the Sombrero's nucleus
differs from Seyfert 1's nuclei: (1) the latter exhibit iron K
emission lines produced in the accretion disc, whose strength clearly
increases with decreasing X-ray luminosity; the average EW of the 6.4
keV emission (for a broad Gaussian fit) is $\sim 300$ eV at an
X-ray luminosity of few$\times
10^{42}$ erg s$^{-1}$ \citep{nan97b}, while our analysis gives an
upper limit of EW$<296$ eV at a much lower luminosity (Sect. 3).  (2)
Seyferts show evidence for X-ray flux variability, with the amplitude
anticorrelated with the source luminosity \citep{nan97a}.  Neither
short term nor long term variability has been detected in or between
our and previous X-ray observations of the Sombrero's nucleus
\citep{fab97,pel02,ter02}.  In the 18ks ACIS-S data we have looked for
variability to take full advantage of the unprecedented angular
resolution.  The pileup does not affect the temporal variation (as we
expect a constant fraction of X-rays migrating from low energies to
higher energies). We have applied a traditional time-binned histogram
analysis and a Bayesian block technique for unbinned data, which was
developed for the $Chandra$ Multi-wavelength Project (Kim et
al. 2003).  In both tests we do not find any statistically significant
variability.  The XMM flux ($F(2-10$ keV)$=1.3\times 10^{-12}$ erg
s$^{-1}$ cm$^{-2}$) is also consistent with that given by the
$Chandra$ snapshot data \citep[$F(2-10$ keV)$=1.6\times 10^{-12}$ erg
s$^{-1}$ cm$^{-2}$,][]{pel02}.  The somewhat larger flux estimated
from the ASCA and BeppoSAX data
is entirely consistent with the much larger aperture used in these
cases. (3) Finally, the small $N_H$ derived from the XMM data is
inconsistent with
the absence of prominent broad line emission 
\citep{pan02}, which contrasts with the predictions of the unification
models of bright Seyfert nuclei \citep{an93}.

\subsection {Radiatively inefficient accretion?}

An alternative possibility is that accretion proceeds at a rate not
far from $\dot M_{Bondi}$ but with a low radiative efficiency, through
an advection dominated optically thin quasi spherical hot accretion
flow \citep[ADAF; e.g.,][]{nar95}. In these models, unlike in the
standard, radiatively efficient, thin disc models, very little of the
gravitational potential energy of the inflowing gas is radiated away
and accretion lacks the cold, optically thick gas required for the
emission of the fluorescent 6.4 keV Fe-K line. The majority of the
observable emission is in the X-ray and in the radio bands. In the
X-rays it comes from thermal bremsstrahlung with a flat power law
($\Gamma < 1.5$ in the $Chandra$ band) or inverse Compton scattering
of soft synchrotron photons by the flow electrons, with a steeper
X-ray spectral shape \citep[e.g., $\Gamma \sim 2.2$ has been used to model
the nucleus of M87,][]{dim03}. In the radio band, synchrotron
emission arises from the strong magnetic field in the inner parts of
the accretion flow.

Detailed ADAF modeling has been applied many times to low luminosity
AGNs and normal ellipticals \citep[e.g.,][]{qua99,dim01,low01}.  The
absence of a fluorescent Fe line from cold material and the observed
X-ray photon index of the Sombrero's nucleus are in accordance with
the ADAF predictions. Also, the radio emission from an ADAF scales
with the SMBH mass and the X-ray luminosity as $L_{15\,GHz}\sim
10^{36} (M_{BH}/10^7M_{\odot}) (L_{2-10{\rm keV}}/10^{40}$erg
s$^{-1})^{0.14}$ erg s$^{-1}$ \citep{yi99}. This relation gives a
predicted $L_{15\,GHz}\sim 1.06\times 10^{38}$ erg s$^{-1}$ for
Sombrero. VLBI measurements \citep{hum84} reveal a flat spectrum core
source
with $L_{15\,GHz}=1.58\times 10^{38}$ erg s$^{-1}$, therefore the
radio--to--X-rays ratio agrees with the predictions of the basic
ADAF. However, an ADAF modeling tailored to the Sombrero' nucleus, and
using a range of $\dot M_{Bondi}=0.01-0.1 M_{\odot}$ yr$^{-1}$ that
includes the range found here, showed that the ADAF synchrotron
emission overestimates significantly (by as much as a factor of $\sim
10$) the observed radio flux \citep{dim01}.

More recent work on radiatively inefficient accretion has revealed
that little mass available at large radii is actually accreted on the
SMBH and most of it is lost to an outflow or circulates in convective
motions \citep{bla99,sto99,ig03}.  If the density is lower in the
inner flow regions, the synchrotron emission in the radio band is
drastically reduced. In the case of Sombrero, strong mass loss with
only a few percent of $\dot M_{Bondi}$ actually accreted on the SMBH
helps reconcile the model with the observations {\it in the radio}
band \citep{dim01}.  However, in these models the radio--to--X-rays
ratio changes with respect to the 'basic' ADAF predictions, in the
sense that outflows or convection suppress the radio emission more
than the X-ray one \citep{quanar99}.  Therefore, if these models are
normalized to match the observed X-ray luminosity of the Sombrero's
nucleus, they will exhibit a deficit of radio emission with respect to
the observations. In this picture, an additional radio source (e.g., a
nuclear jet) could solve the discrepancy.  The presence of jets was
suggested by \citet{dim01}, who in fact best reproduced the flat radio
spectral shape of the Sombrero's nucleus with self-absorbed
synchrotron emission originating from a compact region (while the ADAF
synchrotron radiation has an inverted spectrum up to high radio
frequencies, above which it abruptly drops).

In conclusion, even in the case of low radiative efficiency the mass
accretion rate on the SMBH must be less than $\dot M_{Bondi}$ to match
the observed emission level.  Note that the $Chandra$ upper limits on
the X-ray nuclear luminosities of three ellipticals also require
strong outflows or convection in an ADAF \citep{low01}; and this could
be the case also for the nucleus of M32 \citep{ho03}. Note finally
that a few other low luminosity AGNs exhibit an excess of radio
emission relative to the predictions of low radiative efficiency
models \citep[see][]{qua99,ul01,dim01}.

\subsection{Jet dominated emission?}

Recent radio observations show that a strong jet presence is a common
feature of LINERs \citep{nag01} and recent modeling of low luminosity
AGNs \citep{fal99,yua02} suggests that the higher wavelengths may also
be dominated by the jet emission. In this modeling the accretion flow
can be described either by a standard Shakura-Sunyaev disc or an ADAF,
and a fraction of the accretion flow is advected into a jet, which
represents an important component of the emission spectrum (actually
dominating in low luminosity sources).  In fact, \citet{liv99} have
argued that the \citet{bz77} mechanism might be most relevant for
advection dominated flows.

In these models the synchrotron emission from the base of the jet
produces a ``bump'' of emission from radio to high IR frequencies,
peaking roughly at mid-IR. The flat/inverted radio synchrotron
spectrum comes from further out in the jet. Its optically thin part,
plus self-Comptonized emission and external Compton emission from the
jet, produce the X-ray spectrum.  A flat radio spectrum \citep{hum84}
and a peak in the IR region are features shown by the SED of the
Sombrero's nucleus (Fig.~\ref{sed}). Moreover, such a jet dominated
model has been applied recently to explain the whole spectral energy
distribution of the LINER in the elliptical galaxy IC1459
\citep{fab03}. Similarly to the Sombrero's nucleus, the X-ray
emission of IC1459 is modeled with an unabsorbed power law
of $\Gamma=1.88\pm 0.09$ (using the $Chandra$ ACIS-S data).

It is interesting to note that in this modeling the jets are
by far the dominant sink of power, in the form of kinetic and
internal energy, to the point that $\sim 0.1 L_{acc}$ could be stored
in them. A similar fraction was estimated for the jet power in the
case of the FRI galaxy IC4296, using again ACIS-S data to estimate
$L_{acc}$ plus a direct calculation of the jet kinetic
power, instead of a modeling of the SED \citep{pel03}.

\subsection{Low mass supply?}

Alternatively to the possibilities described in the previous section,
the mass supply to the central SMBH could be much lower than estimated
in Sect. 4, because accretion is not an adiabatic process and perhaps
it is not even steady. This is the case of feedback modulated accretion
models where the interstellar medium is heated by the energy input
from a central source that decreases or stops recursively the
accretion on the central SMBH, giving raise to cycles of
activity. Such a scenario has been argued several times over the last
decade: the ISM could be heated by the impact of collimated outflows
\citep{tab93,bin99,kai03,dim01,dim03} or by inverse Compton scattering
of hard photons \citep{cio01}, both coming from the central
nucleus. Therefore, accretion is only episodic, and between phases of
accretion activity is switched off with a low nuclear luminosity.  In
the \citet{cio01} scenario, at the present time the duty cycle of
activity includes short bright phases followed by prolongued periods
of re--''adjustment'' of the gas surrounding the nucleus, so that only
1/$10^4$ to 1/$10^3$ nuclei in the nearby Universe should be caught in
the bright AGN phase.

Signatures of the central heating include disturbances in the X-ray isophotes
of the gas. In Sombrero, the X-ray isophotes in the circumnuclear
region shown by the ACIS-S image (Fig.~\ref{imachandra}) do not reveal
clear disturbances from nuclear outflows/hard photons heating the ISM
with respect to a round distribution. However, this aspect is
difficult to assess given the crowding of stellar sources close to the
center and it is under study. The same holds for the possible presence
of temperature variations in the circumnuclear region. 

We just note three observational facts that certainly contrast with
the predictions of a global ``cooling flow'' in this bulge: a) the hot
gas temperature shows an increase close to the center, instead of a
smoothly decreasing profile (Fig. 6), even though the errors are large
to establish firmly that central heating is occurring; b) the mass
accretion rate $\dot M_{Bondi}$, although derived for a steady and
adiabatic case that may not apply, is already much less than the
stellar mass loss rate $\dot M_*$ in this bulge [$\dot M_*\sim 0.36$
M$_{\odot}$ yr$^{-1}$ from the prescriptions of the stellar evolution
theory, e.g., inserting the bulge luminosity $L_B=2.4\times 10^{10}
L_{\odot}$ \citep{ho97} in eq. (3) of Ciotti et al. 1991]; c) the
ratio $L_X/L_B$ for the diffuse emission in the bulge ($L_X=9\times
10^{39} $ erg s$^{-1}$ from \citet{fab97}) is far too low for the hot
gas to be in a global inflow \citep{cio91,fab92}. Points a,b,c are
instead consistent with a nuclear induced global degassing of the
bulge, as predicted for low mass systems by the \citet{cio01} heating
scenario.  A detailed analysis and modeling of the status of the hot
gas on the galactic scale is in progress.

\section{Conclusions}

By exploiting XMM's sensitivity 
and  $Chandra$'s high angular resolution 
 we have studied the X-ray properties of
the nucleus of the Sombrero galaxy.  We have then examined various
possibilities for the observed low level of activity shown by this
well known LINER, and these can be summarized as follows:

$\bullet$ The presence of heavy obscuration is ruled out by the observed 
properties of the X-ray spectrum.

$\bullet$ The nuclear bolometric luminosity is at least $\sim 200$ times 
lower than expected if the central SMBH accreted matter at the rate
predicted by the steady, spherical and adiabatic Bondi accretion
theory, with the high radiative efficiency of a standard accretion disc.
Therefore, the final stages of accretion are not radiatively efficient and/or
the mass supply on the SMBH is far less.

$\bullet$ In a radiatively inefficient accretion scenario, where the
rate of accreting mass is close to the Bondi estimate, the
model luminosity is higher than observed. Including outflows or
convection can fix this problem, but then, if the observed X-ray emission is
accounted for, the radio luminosity is underreproduced.  It could then
come from nuclear jets.

$\bullet$ A jet dominated emission model, recently applied to a few
other low luminosity AGNs, could perhaps explain the observed spectral
energy distribution. If this model applies, a large fraction of the
accretion luminosity $L_{acc}$ could be stored in the form of jet power.

$\bullet$ Another possibility is that accretion is not steady and
adiabatic, therefore the Bondi theory does not apply. That is, due to
the presence of a central heating source, far less mass than predicted
by the Bondi accretion rate actually reaches the SMBH. Observed
signatures of a central heating could be an increase in the
temperature profile close to the nucleus and a low hot gas content for 
the bulge.

$\bullet$ In addition to the absence of a standard accretion disc,
other important differences with the properties of luminous type 1
AGNs include the lack of flux variability and Fe-K emission. Also,
this LINER 2 shows no strong absorption therefore the orientation
dependent unified scheme devised for bright AGNs does not apply to it.

\acknowledgments
We thank A. Zezas for help with the data analysis and discussions.
This work was supported under the NASA XMM GO program (PI G. Fabbiano), and
the Chandra X-ray Center contract (NAS 8-39073). S.P. acknowledges
funding from ASI (ASI contracts IR/037/01 and IR/063/02) and MURST.

\clearpage



\clearpage 

\clearpage 

\clearpage 

\clearpage 

\clearpage 

\clearpage 

\clearpage


\begin{deluxetable}{lrrcc}
\tabletypesize{\small}
\tablecaption{XMM-$Newton$ EPIC and $Chandra$ ACIS observation log \label{obslog}}
\tablewidth{0pt}
\tablehead{
\colhead{Instrument} &
\colhead{Obs ID} &
\colhead{Date} &
\colhead{Filter/Grating} &
\colhead{Performed duration (ks)}
}
\startdata
EPIC MOS1&0084030101&December 28, 2001&Thin&42.90\\
EPIC MOS2&0084030101&December 28, 2001&Thin&42.90\\
EPIC pn&0084030101&December 28, 2001&Thin&18.80\\
ACIS-S&1586&May 31, 2001&None&18.75\\
\enddata

\end{deluxetable}

\clearpage

\begin{deluxetable}{lrr}
\tabletypesize{\small} \tablecaption{Results of the EPIC XMM-$Newton$
spectral analysis of the Sombrero nucleus: absorbed power law
model. Fluxes are observed and corrected for the EEF, 
luminosities are intrinsic. \label{specfit1}} \tablewidth{0pt}
\tablehead{ \colhead{ } & \colhead{$20^{\prime\prime}$ extraction
radius} & \colhead{$7^{\prime\prime}$ extraction radius} } \startdata
$\chi^2/dof$&556.9/527&330.2/320\\ Intrinsic N$_H$
(cm$^{-2}$)&$1.3\pm0.1\times10^{21}$&$1.8\pm0.1\times10^{21}$\\ Photon
Index&$1.89\pm0.03$&$1.88_{-0.04}^{+0.05}$\\ 0.5-2 keV flux (erg
cm$^{-2}$
s$^{-1}$)&$6.4\pm0.3\times10^{-13}$&$5.5_{-0.3}^{+0.4}\times10^{-13}$\\
2-10 keV flux (erg cm$^{-2}$
s$^{-1}$)&$1.3\pm0.1\times10^{-12}$&$1.3\pm0.1\times10^{-12}$\\ 0.5-2
keV luminosity (erg
s$^{-1}$)&$6.8\pm0.3\times10^{39}$&$5.7\pm0.4\times10^{39}$\\ 2-10 keV
luminosity (erg
s$^{-1}$)&$1.5\pm0.1\times10^{40}$&$1.4\pm0.1\times10^{40}$\\ \enddata

\end{deluxetable}

\begin{deluxetable}{lrr}
\tabletypesize{\small}
\tablecaption{Results of the EPIC XMM-$Newton$ spectral analysis of the Sombrero
nucleus: absorbed power law plus thermal model. Fluxes are observed and corrected for the EEF, luminosities are intrinsic. \label{specfit2}}
\tablewidth{0pt}
\tablehead{
\colhead{ } &
\colhead{$20^{\prime\prime}$ extraction radius} &
\colhead{$7^{\prime\prime}$ extraction radius}
}
\startdata
$\chi^2/dof$&537.3/525&326/318\\
kT (keV)&$0.61_{-0.12}^{+0.11}$&$0.67_{-0.21}^{+0.27}$\\
Intrinsic N$_H$ (cm$^{-2}$)&$1.4\pm0.1\times10^{21}$&$1.9_{-0.2}^{+0.1}\times10^{21}$\\
Photon Index&$1.86\pm0.03$&$1.89_{-0.07}^{+0.02}$\\
0.5-2 keV flux (erg cm$^{-2}$ s$^{-1}$) (power law)&$6.0\pm0.3\times10^{-13}$&$5.2\pm0.4\times10^{-13}$\\
0.5-2 keV flux (erg cm$^{-2}$ s$^{-1}$) (thermal comp.)&$2.6_{-0.8}^{+0.9}\times10^{-14}$&$8.8\pm8.8\times10^{-15}$\\
0.5-2 keV luminosity (erg s$^{-1}$) (power law)&$6.4\pm0.3\times10^{39}$&$5.5\pm0.4\times10^{39}$\\
0.5-2 keV luminosity (erg s$^{-1}$) (thermal comp.)&$2.7_{-0.9}^{+1.0}\times10^{38}$&$9.3\pm9.3\times10^{37}$\\
\enddata

\end{deluxetable}

\end{document}